\documentstyle[preprint,aps,epsf]{revtex}
\tighten
\begin{document}
\draft
\title{Effective action for
\\bubble nucleation rates}
\author{Ian G. Moss and Wade Naylor\thanks{Current address: Department of Earth \& 
Space Science, Osaka University, OSAKA 560-0043, Japan}}
\address{Department of Physics, University of Newcastle Upon Tyne, NE1
7RU U.K.}
\date{June 2001}
\maketitle
\begin{abstract}
We develop a method to calculate the prefactor in the expression for the 
bubble nucleation rate. A fermion with Yukawa coupling is considered where 
a step potential can be used as a good approximation in the thin wall limit.
Corrections due to thicker walls are investigated by perturbing about the thin wall
case. We derive the thermal one loop effective action, calculating it numerically, 
and find that the prefactor in the nucleation rate can both suppress and enhance, 
for a given temperature, when the usual renormalisation conditions are applied to 
the effective potential.
\end{abstract}
\pacs{Pacs numbers: 03.70.+k, 98.80.Cq}
\narrowtext
\section{Introduction}
Bubble nucleation can occur in a first order phase transition from the
false to the true vacuum. The bubble nucleation rate per unit 
volume per unit time, written in the language of Coleman\cite{coleman}, 
is $\Gamma/V = A\,e^{-B}$, where B is the classical Euclidean action of 
the bubble and A is the one loop contribution including zero and negative modes. 

The purpose of this paper is to present a method that enables one to calculate 
the nucleation rate in the thin wall limit. Techniques such as the derivative expansion 
break down for this type of background\cite{krip}. The thin wall limit is a possible 
scenario in electroweak theory with multiple Higgs fields\cite{zarikas}.
It is also useful as a possible explanation for the generation of the baryon asymmetry 
we observe today\cite{andy}. In the interests of brevity we focus on just the fermion
fields. However, the general discussion includes scalar and spinor fields for completeness.
The calculation of other fields, using the method presented, is currently under way
\cite{naylor}.

Previous work on nucleation rates includes an analysis of the prefactor for scalar 
fields coupled to fermions by Gleiser, Marques and Ramos\cite{gleiser}, who computed the 
determinant using the derivative expansion. Issues regarding which loop corrections should be 
included in the bounce solution were investigated.
Garriga\cite{garriga} calculated the determinantal prefactor analytically, assuming the free 
energy can be approximated by that of a massless field living on the surface of a membrane. 
This was for a scalar field in the thin wall approximation (corrections due to thicker walls 
were also studied) at  finite temperature, where enhancement of the nucleation rate was 
found. Kripfganz, Laser and Schmidt \cite{krip} computed the prefactor for the one loop Higgs 
fluctuations at the electroweak phase transition, using the derivative expansion as an 
approximation.

A particularly useful numerical method uses a theorem on functional determinants that can 
be found in Coleman's work \cite{detcole}. 
Baacke and Kiselev \cite{baacke1} develop an `exact' numerical scheme to work out the 
one loop corrections for a scalar field at finite temperature, using Coleman's theorem. 
Thick and thin walls were considered but an infinitely thin wall 
($\alpha=1$ in their notation) was not. Hence, there is no direct method of comparison with 
\cite{garriga}. However, they also found an enhancement of the nucleation rate (depending
on the choice of renormalisation scheme).
Baacke \cite{baacke2} then studied vector bosons in the 't Hooft-Feynman gauge, 
generalising Coleman's theorem to the matrix case. 
The results were compared to those in \cite{krip}, where there was good agreement for 
small (thick wall) bubbles and strong deviations between the results for large (thin wall) 
bubbles, as expected. The nucleation rate was suppressed for this case.
Also, Baacke and S\"urig \cite{baacke3} calculated the fermionic fluctuation determinant 
using Coleman's theorem and a gradient expansion for comparison. For an optimal choice 
of renormalisation, they found that the rate was enhanced at the electroweak  temperature.

Brahm and Lee\cite{lee} used a different numerical procedure based on the phase shift method
and similar to the one we shall employ (using some old results due to Schwinger 
\cite{schwing}). They computed the prefactor for scalar fields at finite temperature for the thin 
wall limit, based on the assumption that the surface free energy is equal to that of a domain 
wall (thick walls were also considered). The  WKB approximation was used for the high 
energy modes to improve convergence of the exact result. The results were compared with 
the effective potential and derivative expansion approximations (see references therein).  

Further work has been done by M\"unster and Rotsch \cite{munster}, who calculated the 
prefactor in the nucleation rate for a scalar field using a P{\"o}schl-Teller potential
and heat kernel coefficients to remove the ultraviolet divergences of the theory.
M\"unster, Strumia and Tetradis \cite{tetrad} have compared this work to an entirely 
different method (not relying on saddle point evaluation) that uses coarse graining and 
renormalisation group techniques. They find  good agreement between the two in the region
of validity. Further work has been done using this coarse graining method and 
we refer the reader to \cite{tetrad} and the references therein.

In our method we shall use phase shifts and relate the prefactor to the heat kernel. 
The theory is regulated by subtracting off the relevant heat kernel coefficients. 
We consider fermions with a Yukawa coupling and a step function profile to begin with.
Using the step function gives an exact analytic expression for the phase shift,
making the calculation more manageable. For the fermion case, the thin wall limit 
introduces problems of its own because the derivative of the mass term leads to a delta 
function in the potential. As far as we are aware, this is the first time this specific case has 
been investigated, although the method used in \cite{baacke2} or \cite{lee} could be applied.
Results are at finite temperature (not only in the high temperature limit) and the method 
can be extended to the zero temperature case. Also, the technique involves a simple 
regularisation step, unlike methods based on Coleman's theorem which require evaluating
uniform asymptotic expansions of the relevant field equations.

The eigenvalues in the determinant are found using partial wave analysis and 
phase shifts, with the eigenmodes discretised by putting them in a sphere of large radius
$\Omega$ that we let tend to infinity (not to be confused with the bubble wall radius at $R$).
Thus, we must impose boundary conditions on the fields. In the case of fermions, the correct 
eigenvalue problem requires mixed boundary conditions (see appendix A). 

The paper is organised as follows. In Sec. II A we relate the phase shift to the 
heat kernel and zeta function. In Sec. II B  we calculate the thermal effective action leading
to an expression for the prefactor in terms of the phase shift. 
In Sec. III we discuss how to consider corrections from thicker walls. In Sec. IV results 
are presented and in Sec. V we draw conclusions. In Sec. VI an appendix in three
sections is given, with details on the calculation of the fermion phase shift, numerical zeta 
function regularisation  and renormalisation respectively.

\section{Heat kernels and phase shifts}
\subsection{Nucleation \& regularisation}
We begin with the nucleation rates for the decay of a false vacuum at
finite temperature due to an instanton $\phi_{\,bubble}$. Any field that
acquires a mass on the instanton background can contribute to the
prefactor. In three dimensions\cite{linde} (see also comments 
in\cite{gleiser} \& \cite{lee}),
\begin{equation}
A=T \left(\frac{B}{2\pi}\right)^{3/2}
\prod_{fields}\left|\frac{det'[-\nabla^2+m^2(\phi_{\,bubble})]}
{det[-\nabla^2+m^2(\phi_{\,sym})]}\right|^{-1/2}.
\end{equation}
Three zero eigenvalues, arising from breaking the Poincare symmetry,
each contribute $(B/2\pi)^{1/2}$ to the total and these are omitted
from the scalar determinant, as indicated by the prime.

In the thin wall limit we assume that $\phi_{\,bubble}$ has a bubble
wall at some radius $R$, such that $\phi$ takes the false vacuum value
at radii $r>R$ and the true vacuum value at radii $r<R$, with a narrow
transition region near $r=R$. For scalar bosons and fermions 
(and if calculated, the vector bosons), 
the relevant mass terms vanish in the false vacuum and are non-zero in the
true vacuum, leading to a step function profile,
\begin{equation}\label{step}
m(r)=\cases{m&$r<R$\cr0&$r>R$\cr}.
\end{equation}
In the same limit, the scalar Higgs field masses differ little for large and
small radii. This suggests that the Higgs contribution to the
prefactor is smaller than the fermion contribution (and also any other 
fields). We will consider the accuracy of the thin-wall
approximation later.

The eigenvalues in the determinant can be found by using a partial wave
analysis and phase shifts\cite{schiff}. We first discretise the
eigenmodes by putting them in a sphere of large radius $\Omega$. After
separating the eigenmodes into radial functions and spherical
harmonics, the radial parts asymptotically approach trigonometric
functions of $kr+\phi$, where $\phi$ is a constant phase depending on
$k$ and the angular momentum $l$. On the boundary,
\begin{equation}\label{pert}
k_{n}\Omega\approx n\pi - \phi.
\end{equation}
In the false vacuum, the potential is zero and we label the free
eigenvalues $k_n^{(0)}$
\begin{equation}\label{free}
k_{n}^{(0)}\Omega\approx n\pi -\phi^{(0)}.
\end{equation}
On letting $\Omega\rightarrow\infty$ (continuum limit), the above
equations (\ref{pert}) and (\ref{free}) imply a relationship between
the phase shift $\delta_l(k)=\phi-\phi^{(0)}$, the density of states $g_l(k)$ and 
$g_l^{(0)}(k)$,\cite{schwing},
\begin{equation}\label{dens}
g_l(k)=g_l^{(0)}(k) +\frac{1}{\pi}\frac{d\delta_l(k)}{dk},
\end{equation}
for the radial modes.

The difference between the heat kernels for the instanton and the true
vacuum will be
\begin{equation}
\Delta K(t)=\sum_n \left(e^{-k_n^2t}-e^{-k_n^{(0)2}t}\right)
\end{equation}
Using the density of states factor $g_l(k)$ and the degeneracy factor
$\chi_l$,
\begin{equation}
\Delta K(t) =
\int_{0}^{\infty}dk\:e^{-k^2t}\sum_l\chi_l(g_{l}(k)-g_{l}^{(0)}(k)).
\end{equation}
Substituting (\ref{dens}) into the above equation and integrating by parts we
obtain
\begin{equation}\label{ker}
\Delta K(t) = \frac{2}{\pi} \int_{0}^{\infty}dk\:e^{-k^2t}\,kt
\sum_l\:\chi_l\delta_l(k),
\end{equation}
where the degeneracy factor $\chi_l=(2l+1)$ in three dimensions.

The heat kernel can now be used to regularise the determinants
appearing in the prefactor $A$ in the nucleation rate. We define the
generalised $\zeta$ function \cite{dowker} by
\begin{equation}
\zeta(s)= {1\over \Gamma(s)}
\int_{0}^{\infty}\,t^{s-1}\,{\bf tr}\,K(t)\,dt,
\end{equation}
The analytic continuation of $\zeta(s)$ then gives
\begin{equation}
\log A=\sum_{fields}(\pm)\Delta W
\end{equation}
where we take $\pm$ for scalar and spinor fields respectively, and
\begin{equation}
\label{forapen}
\Delta W=\case1/2\zeta'(0)+\case1/2\zeta(0)\log\mu^2
\end{equation}
where $\mu$ is the renormalisation scale.

For numerical work, the analytic continuation can best be performed by
subtracting terms from the heat kernel. As $t\to 0$, the heat kernel in
$d+1$ dimensions has the asymptotic expansion\cite{mossbook}
\begin{equation}\label{kerexp}
K(t) \sim t^{-(d+1)/2}\,\sum_{n=0}\,B_{n}t^{n}\,.
\end{equation}
The leading terms, which cause the poles in the $\zeta$ function, can
be removed by replacing the sum over phase shifts in equation (\ref{ker}) by
\begin{equation}\label{reg}
\sum_l\,\chi_l\bar{\delta}_{l}=
\sum_l\,\chi_l\delta_{l} - \frac{\pi
B_{1}\,k^{d-1}}{\Gamma(\frac{d+1}{2})}
- \frac{\pi B_{3/2}\,k^{d-2}}{\Gamma(\frac{d}{2})}
- \frac{\pi B_{2}\,k^{d-3}}{\Gamma(\frac{d-1}{2})},
\end{equation}
where the $B_0$ coefficient cancels because it is equal to the free heat kernel $K^{(0)}$.
An infrared cutoff $M_{IR}$ must also be included, noting (see Appendix B) that
the dependence on $M_{IR}$ is illusory given that changing $M_{IR}$ does
not affect the value of $\Delta W$. The $B_{3/2}$ coefficient only occurs for fermions 
because squaring the Dirac equation leads to a delta function in the potential
(for a step profile). Heat kernel coefficients have been calculated for distributional
backgrounds \cite{distri} and we only quote the result below.

For the step potential, standard expressions for the heat kernel
coefficients give \cite{mossbook} (\& \cite{distri}):
\begin{eqnarray}
B_{1}&=&-\frac{m^{2}R^{d+1}}{2^{d}(d+1)\Gamma(\frac{d+1}{2})}\label{b1}\,,\\
B_{3/2}&=&\frac{m^{2}R^{d}}{2^{d}\Gamma(\frac{d+1}{2})}\label{b3/2}\,,\\
B_{2}&=&\frac{m^{4}R^{d+1}}{2^{d+1}(d+1)\Gamma(\frac{d+1}{2})}\label{b2}\,.
\end{eqnarray}
For example, the phase shift \cite{schiff} for a scalar boson field is
\begin{equation}\label{delta}
\tan\,\delta_{l}=\frac{k\,J_{l-1/2}(kR)\,J_{l+1/2}(k^{\prime}R) -
k^{\prime}\,J_{l-1/2}(k^{\prime}R)\,J_{l+1/2}(kR)}
		      {k\,N_{l-1/2}(kR)\,J_{l+1/2}(k^{\prime}R) -
		      k^{\prime}\,J_{l-1/2}(k^{\prime}R)\,N_{l+1/2}(kR)}\,,
\end{equation}
where $k^{\prime}=\sqrt{k^{2} -m^{2}}$ and $l=0,1,...$ in three dimensions.
The phase shift for fermions is a rather lengthy calculation that is left 
until Appendix A. 
\subsection{Thermal effective action}
Using the techniques of the last section we are now ready to calculate what
is essentially the difference in the effective action for the true and false vacua
$\Delta W$. We refer the reader to \cite{schof} for a detailed discussion of heat
kernel methods at finite temperature for scalar and spinor fields.
The thermal heat kernel $K^{\beta}$ can be expressed as an
infinite sum of zero temperature heat kernels
\begin{equation}
K^{\beta}(t\mid
\tau,x;\,\tau',x')=\sum_{n=-\infty}^{\infty}(\pm)^{n}\,K(t\mid
\tau,x;\,\tau'+\beta n,x')
\end{equation}
(where $\pm$ is for scalar and spinor fields respectively) and for an
ultrastatic spacetime, the heat kernel
can be factorised into temporal and spacial parts giving
\begin{equation}
K(t\mid \tau,x;\tau',x')=\frac{1}{\sqrt{4\pi
t}}\,e^{-\frac{(\tau-\tau')^{2}}{4t}}K^{(3)}(t\mid x,x').
\end{equation}
It is then possible to show using the above relations that,
\begin{equation}\label{spacial}
K^{\beta}(t)=\frac{\beta}{\sqrt{4\pi
t}}\,\sum_{n=-\infty}^{\infty}\,e^{-\frac{n^{2}\beta^{2}}{4t}} K^{(3)}(t).
\end{equation}
$\Delta W$ is related to the thermal heat kernel $\Delta K^{\beta}(t)$ for
scalar fields by
\begin{equation}
\Delta W=-\frac{1}{2}\int_{0}^{\infty}\frac{dt}{t}\,{\bf
tr}\Delta K^{\beta}(t).
\end{equation}
Substituting (\ref{spacial}) into the above equation, we obtain for scalars,
\begin{equation}\label{frbs}
\Delta W=-\frac{\beta}{2}\int_{0}^{\infty}\frac{dt}{t}\,
\sum_{n=-\infty}^{\infty}e^{-\frac{\beta^{2}}{4t}n^{2}}\frac{1}{\sqrt{4\pi t}}
		  \frac{2}{\pi} \int_{0}^{\infty}dk\:e^{-k^2t}
\,k\,t \sum_{l=0}^{\infty}(2l+1)\:\delta_{l}(k).
\end{equation}
Then, using the fact that
\begin{equation}\label{integral}
\int_{0}^{\infty}t^{-\frac{1}{2}}e^{-\frac{\beta^{2}}{4t}n^{2}-k^{2}t}dt=
\frac{\sqrt{\pi}}{k}e^{-n\beta k},
\end{equation} we get
\begin{equation}
\Delta W=-\frac{\beta}{2\pi}\int_{0}^{\infty}dk
\sum_{l=0}^{\infty}(2l+1)\:{\delta}_{l}(k)
		 -\frac{\beta}{\pi}\int_{0}^{\infty}dk
\sum_{l=0}^{\infty}(2l+1)\:\delta_{l}(k)\sum_{n=1}^{\infty}e^{-n\beta k}.
\end{equation}
The sum over $n$ is standard, leading to the result
\begin{equation}\label{bosew}
\Delta W=-\frac{\beta}{2\pi}\int_{0}^{\infty}dk
\sum_{l=0}^{\infty}(2l+1)\:\bar{\delta}_{l}(k)
		 -\frac{\beta}{\pi}\int_{0}^{\infty}dk
\sum_{l=0}^{\infty}(2l+1)\:\frac{\delta_{l}(k)}{e^{\beta k}-1}.
\end{equation}

The first term is the zero point energy, that contains the ultraviolet
divergences of the theory and hence
$\bar{\delta_{l}}(k)$ (see equation (\ref{spacial}) \& (\ref{reg})). The second term
is the temperature dependent part.
The above expression can be derived using the density of states method\cite{lee}. 
Thus, upon regulating the nonthermal part, using zeta function 
regularization and introducing a mass that we let tend to zero at the end
of the calculation (see Appendix B), we have
\begin{equation}\label{boswn}
\Delta W^{N}=-\frac{\beta}{2\pi} \int_{0}^{\infty}dk
\left(\:\sum_{l=0}^{\infty}(2l+1)\delta_{l}(k)-2\sqrt{\pi} B_{1}k
			-\frac{\sqrt{\pi}
			B_{2}}{\sqrt{(k^{2}+M^{2}_{IR})}}\right)
+ \frac{\beta B_{2}}{2\sqrt{4\pi}}\,logM^{2}_{IR},
\end{equation}
\begin{equation}\label{boswt}
\Delta W^{T}=-\frac{\beta}{\pi}\int_{0}^{\infty}dk
\sum_{l=0}^{\infty}(2l+1)\:\frac{\delta_{l}(k)}{e^{\beta k}-1},
\end{equation}
where $\Delta W=\Delta W^{N}+\Delta W^{T}$ is the thermal effective
action and we are working explicitly in three dimensions. (Note that $\Delta W$ is
independent of $M^{2}_{IR}$ and for the scalar boson, $B_{3/2}$ is zero.)

For fermions the spinor effective action $\Delta W_{(1/2)}$ is related
to the heat kernel $\Delta K^{\beta}_{(1/2)}(t)$ by (twice the scalar result for
massive fermions and also a colour factor of three if the top quark is considered)
\begin{equation}
\Delta W_{(1/2)}=\int_{0}^{\infty}
\frac{dt}{t}\,$tr$K^{\beta}_{(1/2)}(t).
\end{equation}
Thus,
\begin{equation}
\Delta W_{(1/2)}=4\beta\int_{0}^{\infty}\frac{dt}{t}\,
\sum_{n=-\infty}^{\infty}(-1)^{n}e^{-\frac{\beta^{2}}{4t}n^{2}}\frac{1}{\sqrt{4\pi t}}
		  \frac{2}{\pi} \int_{0}^{\infty}dk
\:e^{-k^2t}\,k\,t \sum_{j=1/2}^{\infty}2(2j+1)\:\delta_{f}(k),
\end{equation}
where $\delta_{f}=\delta_{+}+\delta_{-}$ (see appendix) and $j= 1/2,3/2,...$ 
in three dimensions (the factor of four comes from the trace over spinor indices).
Applying (\ref{integral}) gives
\begin{equation}
\Delta W_{(1/2)}=\frac{4\beta}{\pi}\int_{0}^{\infty}dk
\sum_{j=1/2}^{\infty}2(2j+1)\:\delta_{f}(k)
-\frac{8\beta}{\pi}\int_{0}^{\infty}dk
\sum_{j=1/2}^{\infty}2(2j+1)\:\delta_{f}(k)\sum_{n=1}^{\infty}
(-1)^{n}e^{-n\beta k}.
\end{equation}
Then, summing over n we have
\begin{equation}
\Delta W_{(1/2)}=\frac{4\beta}{\pi}\int_{0}^{\infty}dk
\sum_{j=1/2}^{\infty}2(2j+1)\:\bar{\delta}_{f}(k)
-\frac{8\beta}{\pi}\int_{0}^{\infty}dk
\sum_{j=1/2}^{\infty}2(2j+1)\:\frac{\delta_{f}(k)}{e^{\beta k}+1}.
\end{equation}

Of course, we could have guessed this result from looking at
(\ref{bosew}), taking into account the properties of spinors. 
It is fairly simple to derive the above equation using the density 
of states in the same way as in \cite{lee}, using the spinor
phase shift. Thus,
\begin{eqnarray}\label{ferwn}
\Delta W^{N}_{(1/2)}&=&\frac{4\beta}{\pi} \int_{0}^{\infty}dk
\left(\:\sum_{j=1/2}^{\infty}2(2j+1)\:\delta_{f}(k) -2\sqrt{\pi} B_{1}k
-\pi B_{3/2} -\frac{\sqrt{\pi} B_{2}}{\sqrt{(k^{2}+M^{2}_{IR})}}\right)\nonumber\\
&&- \frac{2\beta B_{2}}{\sqrt{\pi}}\,logM^{2}_{IR},
\end{eqnarray}
\begin{equation}\label{ferwt}
\Delta W^{T}_{(1/2)}=-\frac{8\beta}{\pi}\int_{0}^{\infty}dk
\sum_{j=1/2}^{\infty}2(2j+1)\:\frac{\delta_{f}(k)}{e^{\beta k}+1}.
\end{equation}
Note the opposite sign in the zero point energy (nonthermal) contribution to
the one loop effective action as compared to the scalar case equation (\ref{boswn}).
\section{Thicker walls}
The phase shift method works equally well for any bubble profile, although
a differential equation must be solved numerically to find the phase shift.
In the general case, one could consider the difference between the phase shift
and thin wall phase shift and then add this correction onto the effective action 
numerically. Alternatively, it is possible to estimate the corrections due to the
non-zero thickness of the wall by perturbing about the thin wall case. For example, 
consider the scalar boson, for which we must solve
\begin{equation}
-r^{-2}(r^{2} u^{\prime})^{\prime} +m^{2}u +\frac{l(l+1)}{r^2}u -k^{2}u = -Vu,
\end{equation}
where $m$ is given by equation (\ref{step}), $V$ is the correction due to
a thicker wall and we define $R$ such that $\int^{\infty}_{0}r^{2}V(r)dr=0$. 
The Green's function is
\begin{equation}
	G(r,r^{\prime})=-\cases{kA^{-1}u_{1}(r)u_{2}(r^{\prime})&$r<r^{\prime}$
	                 \cr kA^{-1}u_{2}(r)u_{1}(r^{\prime})&$r>r^{\prime}$\cr}
\end{equation}
and
\begin{eqnarray}
	 & u_{1}(r)=\cases{ j_{l}(k^{\prime}r)& $r<R$\cr Aj_{l}(kr)-Bn_{l}(kr)& $r>R$\cr};\;
     & u_{2}(r)=\cases{ Cj_{l}(kr^{\prime})+Dn_{l}(kr^{\prime})& $r<R$\cr n_{l}(kr)& 
     $r>R$\cr},
\end{eqnarray}
where $k^{\prime}=\sqrt{k^{2}-M^{2}}$ and 
$j_{l}(z)=\sqrt{\frac{\pi}{2}}z^{(1-d)/2}J_{l+(d-1)/2}(z)$ in $d+1$ 
dimensions.

One then imposes $u \propto j_{l}$ as $r\rightarrow 0$ and $u \propto 
Aj_{l}-B^{\prime}n_{l}$ as $r \rightarrow \infty$. Then, the solution is
\begin{equation}
u=u_{1} - \int^{\infty}_{0}G(r,r^{\prime})V(r^{\prime})u(r^{\prime})\,r^{\prime 2}\,dr^{\prime}.	
\end{equation}
Therefore, as $r\rightarrow\infty$
\begin{equation}
u\rightarrow u_{1} + 
kA^{-1}u_{2}\int^{\infty}_{0}V(r^{\prime})u^{2}_{1}(r^{\prime})\,r^{\prime 2}\,dr^{\prime}.	
\end{equation}

From this it is possible to show that the correction to the phase shift is
\begin{equation}
\delta^{(1)}=-\frac{k}{A^{2}+B^{2}}
\int^{\infty}_{0}V(r)u^{2}_{1}(r)\,r^{2}\,dr.
\end{equation}
Then, assuming that the Bessel functions 
change little as $r$ varies over the bubble wall, they can be Taylor 
expanded about $R$, giving
\begin{equation}\label{thick}
\delta^{(1)}\approx-\frac{2kk^{\prime}}{A^{2}+B^{2}}\;j_{l}(k^{\prime}R)j_{l}^{\prime}
(k^{\prime}R)\int^{R}_{0}V(r)\,r^{3}\,dr,	
\end{equation}
where the continuity of $u_{1}$ at the bubble wall has been used. For the scalar case, 
\begin{equation}
B=-kR^{2}(k\,j_{l-1}(kR)\,j_{l}(k^{\prime}R) -k^{\prime}\,j_{l}(kR)\,j_{l-1}(k^{\prime}R))
\end{equation}
and
\begin{equation}
A=-kR^{2}(k\,n_{l-1}(kR)\,j_{l}(k^{\prime}R) -k^{\prime}\,n_{l}(kR)\,j_{l-1}(k^{\prime}R)).
\end{equation}
All the terms outside the integral in equation (\ref{thick}) are numerical factors. For reasons 
discussed in the conclusion we only add the correction onto the thermal part of the effective 
action, giving
\begin{equation}
\Delta W^{(1)}=\alpha\frac{\beta}{R^3}\int_0^R V(r)r^3\,dr
\end{equation}
where $\alpha$ is given by
\begin{equation}
\alpha=-\frac{1}{\pi}\int_{0}^{\infty}dz
\sum_{l=0}^{\infty}\:\frac{(2l+1)}{e^{\beta z/R}-1}\frac{2zz^{\prime}}{A^{2}+B^{2}}\;
j_{l}(z^{\prime})j_{l}^{\prime}(z^{\prime}).
\end{equation}
If one assumes a `$tanh$' like potential $m^2(r)$ for the thicker wall, then 
$V=m^2(r)-m^2\approx -m^2/2$ at the bubble wall radius $R$. Therefore, in terms of the width $w$ 
of the bubble wall one obtains the approximate result 
$\Delta W^{(1)}\approx-\alpha\beta m^2 w/2$. We have calculated $\alpha$ numerically and found 
that it is of order $1$ for a range of values of $\beta/R$.

\section{Results}
The thermal one loop effective action was calculated numerically for 
fermions. The phase shift is substituted into equations (\ref{ferwn}) and (\ref{ferwt}),
where it is convenient to change variables $k \rightarrow z=kR$ for the 
step potential. Then the nonthermal part has only one parameter 
$\eta=m^{2}_{f}\,R^{2}$, where $m_{f}$ is the mass of the fermion. 
The thermal part has parameters $\eta$ and $\beta m_{f}$ (given that $\beta/R=\beta 
m_{f}/\sqrt{\eta}$) as independent variables.

We work with equations (\ref{ferwn}) and (\ref{ferwt}) using a numerical 
package. For each value of $z$ the function is summed over $l$ (or $j$), 
with $l$ increasing up to a given $L$ until the value of the function at $z$ 
converges. The thermal part of the integral converges due to the exponential
damping terms. When considering the nonthermal part of the function, we must 
check that the integrand has the correct $k$ dependence after making the subtraction
of the divergent quantities from the sum over the phase shift. This is a 
good check verifying that the heat kernel coefficients are correct.

We integrate up to $Z$ chosen to obtain the required accuracy (all results are 
accurate to 1\%). For large values of $\eta$ (and small $\beta m_{f}$), larger 
values of $Z$ and $L$ are needed to give convergence. 
The $arctan$ function (from the phase shift) has problems with branches 
for large values of $\eta$ (whenever $\delta$ hits $\pm \pi$), requiring 
numerical glueing of the phase shift. 

\begin{figure}
\begin{center}
\leavevmode
\epsfxsize=20pc
\epsffile{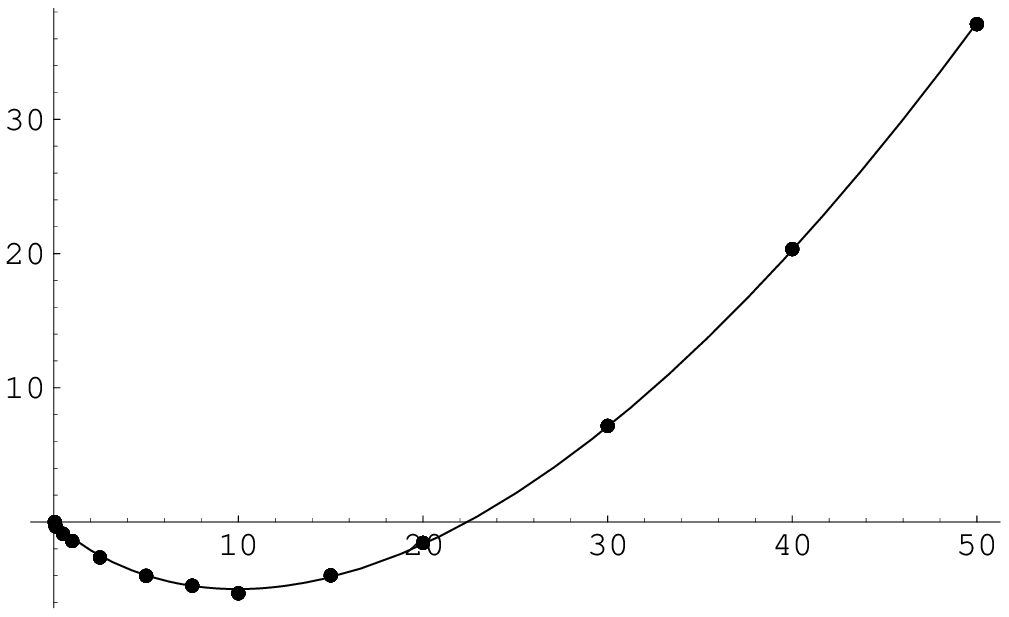}
\end{center}

\caption{Fit of the nonthermal part of the one loop contribution to the fermion 
effective action $\Delta W^N_{(1/2)}$, plotted
against $\eta$.}
\label{fig1}
\end{figure}

The nonthermal part of the fermion effective action can be written $\beta m_{f} F(\eta)$. 
Numerically, $F(\eta)$ fits well to a power law dependence on $\eta$ (see Fig. \ref{fig1}), 
giving, in the original variables,
\begin{equation}
\Delta W^{N}_{(1/2)} =-1.51\beta m^{3}_{f}R^{2} +0.32\beta m^{4}_{f}R^{3} 
\label{fit}
\end{equation}
where we have set the renormalisation scale $\mu=M_{IR}=m_{f}$ as explained in Appendix C. 
The nonthermal part is plotted in Fig. \ref{fig2}. The full one loop 
effective action for fermions plotted against $\eta$ is in Fig. \ref{fig3} for 
various values of the parameter $\beta m_{f}$.

\section{Conclusion}

We have presented a simple method to compute the prefactor in the 
expression for the bubble nucleation rate, applying this to infinitely thin walls. 
Analytic corrections due to thicker walls were considered by perturbing 
about the thin wall case. It is easy to see that these corrections do not effect the $B_{1}$ heat 
kernel coefficient. Using the arguments from Appendix C,  where it was shown that we only 
need to consider the renormalisation of the thin wall bubble, one can ignore the correction from 
$B_{2}$. Thus, only adding corrections to the thermal part of the effective action should be a good
approximation.

In the  context of the electroweak phase transition, we would also like to examine the 
vector bosons, using the method presented. This requires vector spherical harmonics and 
the relevant boundary conditions to calculate the phase shift. Then a full treatment of all the
particle species at the phase transition can be worked out in the thin wall approximation, 
including analytic expressions for corrections due to thicker walls.

The fermion contribution generally enhances the rate, but for large $\beta m_f$ 
(when $\mu=m_f$) it does not and becomes suppressive.
In fact, at $\beta m_f=5.0$ (see figure \ref{fig3}), the sign of the effective action changes for 
various values of $\eta=m_{f}R$ (for a fixed bubble wall radius). 
Choosing $\mu=m_t$ and $m_f=m_t$ (the mass of the top quark) as in \cite{baacke3}, 
then the log term cancels for the fermion determinant (see equation (\ref{fit})). 
Baacke and S\"urig \cite{baacke3} found a negative contribution (enhance), whereas we find it 
can also be positive (suppress) for a temperature ($\frac{1}{\beta}=35$GeV with 
$\beta m_t=5.0$ and $m_t\approx 175$GeV) lower than the that at the electroweak phase 
transition. 
However, there is no direct method of comparison between the two methods because 
in \cite{baacke3} the thin wall limit was not considered. 

The renormalisation scale can be set by imposing conditions on the effective
potential, calculated for constant background fields \cite{lee}. In Appendix C it was shown that 
these conditions on the effective potential (that include the classical and regularised terms) lead
to the choice $\mu=M_{IR}=m_f$. 

In the case of bubble nucleation, the bubble wall radius is found by extremising an action which 
includes the effective potential. The three dimensional action for the bounce solution is
\begin{equation}
B=4\pi \beta R^2\sigma -\frac{4\pi}{3}\beta R^3\varepsilon,
\end{equation}
where $\sigma$ and $\epsilon$ are corrected surface and volume energy densities respectively. 
The nonthermal part of the effective action (equation (\ref{fit})) has a similar form and can be
absorbed into a redefinition of $\sigma$ and $\epsilon$. The value of $R$ should be presumably 
chosen to coincide with the extrema of the new action. Then, the nucleation rate is determined 
as a function of temperature by $\Delta W^T$.

The method described can also be used to calculate the one loop effective action in general, 
but in this case the $B_0$ heat kernel coefficient must be renormalised away in the vacuum 
energy. The numerical procedure is simple and efficient. 

\acknowledgements 
W.N. would like to thank JSPS for Postdoctoral Fellowship for 
Foreign Researchers No. P01773, where this work was revised and completed.
 
\section{Appendix}
\subsection{Fermion phase shift}
Here we present the calculation of the spinor phase shift.
One must use the Dirac equation separated into radial and angular coordinates and also be 
careful in the way one works out the eigenvalues of the problem.
Starting with 
\begin{equation}
(i\gamma \cdot \delta-m)\psi_{+}=\lambda \psi_{-},
\end{equation}
\begin{equation}
(i\gamma \cdot \delta+m)\psi_{-}=-\lambda \psi_{+},
\end{equation}
so that upon squaring the Dirac equation we get the Klein Gordon equation.
The radial components are then\cite{schiff}
\begin{equation}\label{dirac1}
(E \mp m)g_{\pm} + f^{\prime}_{\pm} +(j+\frac{3}{2})\frac{f_{\pm}}{r}=\pm \lambda g_{\mp},
\end{equation}
\begin{equation}\label{dirac2}
(E \pm m)f_{\pm} - g^{\prime}_{\pm} +(j-\frac{1}{2})\frac{g_{\pm}}{r}=\mp \lambda f_{\mp},
\end{equation}
where 
\begin{equation}
\Psi_{\pm}=\left( \begin{array}{c}
i g_{\pm} Y  \\
f_{\pm} \sigma_{r} Y 
\end{array} \right);\qquad
\sigma_{r}=\left( \begin{array}{cc}
 0&-i \\
i&0
\end{array} \right),
\end{equation}
for $j=l+1/2$. The solutions are spherical Bessel functions
$f_{\pm}=A_{\pm}j_{j+\frac{1}{2}}$ and $g_{\pm}=C_{\pm}j_{j-\frac{1}{2}}$ 
(this can be seen easily by substituting (\ref{dirac1}) into (\ref{dirac2}))
where the constants are as yet undetermined. 
On substituting the power series for the Bessel functions into
(\ref{dirac1}) and (\ref{dirac2}) one finds the relations
\begin{equation}
(E\mp m)C_{\pm} + A_{\pm} k\mp \lambda C_{\mp}=0,
\end{equation}
\begin{equation}
(E\pm m)A_{\pm} + C_{\pm} k\pm \lambda A_{\mp}=0.
\end{equation}

The correct eigenvalue problem requires mixed 
boundary conditions\cite{mossbook} (also for a self-adjoint action)
leading to $f_{+}+g_{+}=f_{-}-g_{-}=0$ as $r\rightarrow \infty$.
It is also possible to show that there is a symmetry among the solutions 
such that $f_{+}\leftrightarrow f_{-}$ and $g_{+}\leftrightarrow -g_{-}$.
For the calculation we set $E=0$ and impose the boundary conditions with
the above symmetries, noting that $m\rightarrow 0$ as $r\rightarrow\infty$.
Using this information, it is possible to show that there are two solutions
\begin{equation}
\tan\,\delta=\frac{B_{+}}{A_{+}},
\end{equation}
for $A_{+}=\pm C_{+}$ and $B_{+}=\pm D_{+}$,
where $B_{\pm}$ and $C_{\pm}$ are the corresponding constants
for $n_{j\pm\frac{1}{2}}$, the irregular solutions, that are spherical Neumann
functions.

Then, matching the wavefunction at the junction and performing
some algebra, we get the phase shifts,
\begin{equation}
\tan\,\delta_{\pm}=\frac{(k\pm m)\,J_{j}(kR)\,J_{j+1}(k^{\prime}R)
 - k^{\prime}\,J_{j}(k^{\prime}R)\,J_{j+1}(kR)}
{(k\pm m)\,N_{j}(kR)\,J_{j+1}(k^{\prime}R)
 - k^{\prime}\,J_{j}(k^{\prime}R)\,N_{j+1}(kR)}.
\end{equation}
where all symbols have their usual meanings as in the rest of the paper. 

The $j=l-1/2$
modes have symmetry such that $f_{\pm} \leftrightarrow g_{\pm}$ changes the boundary 
conditions and equations into the $j=l+1/2$ case. Thus we have the above phase 
shifts with a total degeneracy $2(2j+1)$, where $j= 1/2,3/2,...$. All results can be
generalised to the four dimensional case.

\subsection{Numerical zeta function regularisation}
Consider the finite temperature case. The nonthermal part of the heat kernel is
\begin{equation}
\Delta K_0(t)=\frac{\beta}{\sqrt{(4\pi t)}}\frac{2}{\pi}\int_0^\infty dk\,k\,e^{-(k^2+M^2)t}
\sum_\gamma \delta_\gamma(k)\,t,
\label{conker}
\end{equation}
where all symbols have the same conventions as in the rest of the paper and we have introduced 
a mass term $M$.
The zeta function is then
\begin{equation}
\Delta \zeta_0(s)=\frac{\beta}{\sqrt{(4\pi)}}\frac{2}{\pi}\frac{\Gamma(s+\frac{1}{2})}
{\Gamma(s)}\int_0^\infty dk\,k\,(k^2+M^2)^{-s-\frac{1}{2}}
\sum_\gamma \delta_\gamma(k).
\end{equation}
The integral converges for $s>1$.
In order to apply the zeta function to numerical work, we subtract terms from the 
asymptotic expansion of the heat kernel and define
\begin{equation}
P(s)=\frac{2}{\pi}\frac{\beta}{\sqrt{(4\pi)}}\frac{\Gamma(s+\frac{1}{2})}
{\Gamma(s)}\int_0^\infty dk\,k\,(k^2+M^2)^{-s-\frac{1}{2}}\left\{
\sum_\gamma \delta_\gamma-2\sqrt\pi B_1 k-\pi B_{3/2}-\sqrt\pi\frac{B_2}{k}\right\}.
\label{subtract}
\end{equation}
Performing the integration over $k$, and adding back the subtracted terms, one finds
\begin{equation}
\Delta \zeta_0(s)=P(s)+\frac{\beta}{2\sqrt\pi}M^{2-2s}\frac{B_1}{s-1}+
\frac{\beta}{2\sqrt\pi}\frac{\Gamma(s-\frac{1}{2})}{\Gamma(s)}M^{1-2s}B_{3/2}+
\frac{\beta}{2\sqrt\pi}M^{-2s}B_2.
\end{equation}
This is a regular function at $s=0$.
Taking the derivative with respect to $s$ (for $s=0$) and the limit $M\rightarrow 0$, 
we obtain
\begin{equation}
\label{zetareg}
\Delta \zeta^\prime_0(0)=\frac{\beta}{\pi} \int_{0}^{\infty}dk
              \left(\:\sum_\gamma^{\infty}\delta_{\gamma}
            -2\sqrt{\pi} B_{1}k -\pi B_{3/2} -\frac{\sqrt{\pi}B_{2}}{\sqrt{(k^{2}+M_{IR}^{2})}}
         \right) - \frac{\beta B_{2}}{\sqrt{4\pi}}\,\log M_{IR}^{2},
\end{equation}
where $M_{IR}$ is an infrared cutoff. It is important to note that if one subtracts the
difference between $\Delta \zeta^\prime_0(0)$ for different values of infrared cutoff 
($M_A$ and $M_B$ say) and using the fact that
\begin{equation}	
\int_0^\infty dk \left(\frac{1}{\sqrt{k^2+M_A^2}}-\frac{1}{\sqrt{k^2+M_B^2}}\right)
=\log\frac{M_B}{M_A},
\end{equation}
it is clear that this difference is zero. Thus, equation (\ref{zetareg}) is independent of $M_{IR}$.

\subsection{Renormalisation}
The renormalisation scale $\mu$ represents an arbitrary parameter that can be expressed in terms 
of other constants by imposing conditions on the effective potential (see \cite{berkin} for example). 
We shall show that the usual renormalisation conditions for the effective  potential,
\begin{eqnarray}
V'(\phi_{vac})&=&0, \nonumber\\
V''(\phi_{vac})&=&m_h, 
\label{recon}
\end{eqnarray}
where $\phi_{vac}$ is the classical vacuum expectation value and $m_h$ is the mass of the Higgs 
field, give the optimal choice, removing the logarithmic term.

 
The first two terms in the zero point energy (\ref{fit}) come from the finite, subtracted, part of the 
effective action (see equation (\ref{subtract})). A similar approach was 
used in \cite{baacke3}, where the divergent effective action was regularised by subtracting terms 
proportional to 1st and 2nd powers of the background potential, leaving a finite contribution to the 
fluctuation determinant. The two divergent terms were then regulated by using a momentum 
cutoff and renormalised by applying conditions on the propagators. This procedure is equivalent to 
the subtraction of the heat kernel coefficients up to and including $B_2$ (see equation 
(\ref{subtract})). However,  zeta function regularisation is unusual, as compared with most 
regularisation schemes, in that it does not need infinite counter terms. Thus, in equation 
(\ref{zetareg}) only the last term, containing the logarithm, requires renormalisation. 

The effective potential $V=V_0+V_1$ will be defined
\begin{equation}
V=\int\beta \,d^3x\left[\,\frac{1}{2}(m_h^2+\delta m_h^2)\phi^2-(g+\delta g)\phi^3
+\frac{1}{4}(\lambda+\delta \lambda)\phi^4\right]
-\frac{2\beta B_2}{\sqrt{\pi}}\log\frac{M_{IR}^2}{\mu^2},
\end{equation}
where the terms in the square brackets make up the classical potential (including counterterms), 
the last term is the $\zeta$-regularised 1-loop contribution with
\begin{equation}
B_2=\Delta \zeta_0(0)=\frac{\beta}{(4\pi)^{3/2}}\int d^3x\, \frac{1}{2}\,m_f^4\phi^4(\vec{x}),
\end{equation}
and we have included the renormalisation scale $\mu$ (see equation (\ref{forapen})). 

In our case the renormalisation must be applied to a step function field configuration, but the 
ultraviolet divergences are independent of the background profile. This is intuitively 
obvious because $k\rightarrow \infty$ equates to the short distance behaviour of the theory. 
Thus, the $B_2$ heat kernel coefficient ($\propto \phi^4(\vec{x})$) need only be considered for 
$\phi(\vec{x})=constant$, i.e. the effective potential. 




By applying the renormalisation conditions (\ref{recon}) to the effective potential, it is simple to 
show that the counter terms are given by $\delta m_h^2=\delta g=0$ and
\begin{equation}
\delta \lambda=\frac{m_f^4}{\sqrt{2}\pi^2}\log\frac{M_{IR}^2}{\mu^2}.
\end{equation}

On substituting the above counterterms into the effective potential it is clear that the logarithmic 
term will cancel. 
Naturally, for our numerical procedure we can choose $M_{IR}=m_f$ because it was shown in 
Appendix B that the infrared cutoff was independent of the result (equation (\ref{zetareg})).

With our choice of renormalisation we can compare with a similar work \cite{lee}, where the scalar 
Higgs field was investigated. In the case of the thin wall limit they only considered the surface free 
energy. However, comparing with our zero point surface energy contribution, for a scalar boson 
\cite{wthesis}, we find it is of a similar order of magnitude and also suppresses the nucleation rate. 
This contribution is considerably smaller than that made by a fermion with a Yukawa coupling 
which, generally, enhances the nucleation rate. 


\begin{figure}
\begin{center}
\leavevmode
\epsfxsize=20pc
\epsffile{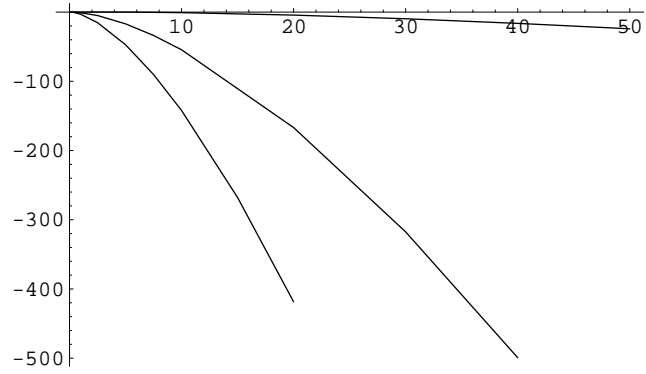}
\end{center}
\caption{The thermal part of the one loop contribution to the fermion effective action 
$\Delta W^T_{(1/2)}$, plotted against $\eta$ for values of $\beta 
m_{f}$ (from bottom to top); 0.5, 1.0 \& 5.0.}
\label{fig2}
\end{figure}

\begin{figure}
\begin{center}
\leavevmode
\epsfxsize=20pc
\epsffile{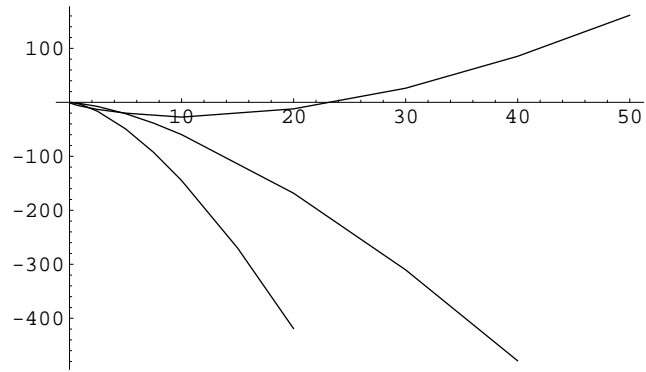}
\end{center}
\caption{The full one loop contribution to the fermion effective action $\Delta 
W_{(1/2)}$, plotted against $\eta$ for values of $\beta 
m_{f}$ (from bottom to top); 0.5, 1.0 \& 5.0.}
\label{fig3}
\end{figure}


\end{document}